\documentclass[aps,pra,noshowpacs,twocolumn,preprintnumbers,superscriptaddress,floatfix]{revtex4}
\usepackage{amsmath,amssymb}

\pdfoutput=1
\usepackage{hyperref}
\usepackage{graphicx}

\begin{document}
\title{Invisibility Cloak Printed on a Photonic Chip}
\newcommand{\SKL}{State Key Laboratory of Advanced Optical Communication Systems and Networks, Institute of Natural Sciences $\&$ Department of Physics and Astronomy, Shanghai Jiao Tong University, Shanghai 200240, China}
\newcommand{\USTC}{Synergetic Innovation Center of Quantum Information and Quantum Physics, University of Science and Technology of China, Hefei, Anhui 230026, China}

\author{Zhen Feng}
\affiliation{\SKL}
\affiliation{\USTC}
\author{Bing-Hong Wu}
\affiliation{\SKL}

\author{Yu-Xi Zhao}
\affiliation{\SKL}

\author{Jun Gao}
\affiliation{\SKL}
\affiliation{\USTC}
\author{Lu-Feng Qiao}
\affiliation{\SKL}
\affiliation{\USTC}
\author{Ai-Lin Yang}
\affiliation{\SKL}
\affiliation{\USTC}
\author{Xiao-Feng Lin}
\affiliation{\SKL}
\affiliation{\USTC}
\author{Xian-Min Jin}
\email{xianmin.jin@sjtu.edu.cn}
\affiliation{\SKL}
\affiliation{\USTC}


\maketitle

\textbf{Invisibility cloak capable of hiding an object can be achieved by properly manipulating electromagnetic field. Such a remarkable ability has been shown in transformation and ray optics. Alternatively, it may be realistic to create a spatial cloak by means of confining electromagnetic field in three-dimensional arrayed waveguides and introducing appropriate collective curvature surrounding an object. We realize the artificial structure in borosilicate by femtosecond laser direct writing, where we prototype up to $5,000$ waveguides to conceal millimeter-scale volume. We characterize the performance of the cloak by normalized cross correlation, tomography analysis and continuous three-dimensional viewing angle scan. Our results show invisibility cloak can be achieved in waveguide optics. Furthermore, directly printed invisibility cloak on a photonic chip may enable controllable study and novel applications in classical and quantum integrated photonics, such as invisualising a coupling or swapping operation with on-chip circuits of their own.}
\section*{Introduction}
Transformation optics is a prevalent approach to modify material parameters, carefully constructing artificial material and manipulating electromagnetic wave \cite{Pendry2006,Leonhardt2006}.  Due to the enormous difficulties in the realization of cloaking, scientists have been seeking various simplified models \cite{Schurig2006,Cai2007,Li2008,Kildishev2008,Leonhardt2009,Chen2010} to approach the prefect cloaking. Among those methods, Schuig \emph{et al.} \cite{Schurig2006} use artificially structured metamaterial to fulfill the dimension-reduced cloaking scheme which could only cloak the light at some specific frequencies and polarizations. To broaden the operating frequency range \cite{Li2008,Kildishev2008,Leonhardt2009}, the carpet cloak is introduced to operate at microwave \cite{Liu2009,Ma2010} and optical \cite{Valentine2009,Gabrielli2009,Ergin2010} frequency bands. Metamaterials are also applied to achieve a skin cloak which can hide volumetric objects at optical frequencies \cite{Ni2015}. In diffusive light scattering medium where light does not conform to the ballistic propagation, broadband macroscopic cloak made of thin shells of metamaterial is realized in three dimensions, according to Fick$^\prime$s equation \cite{Schittny2014}. \par
\begin{figure*}[!t]
\includegraphics[width=0.98\textwidth]{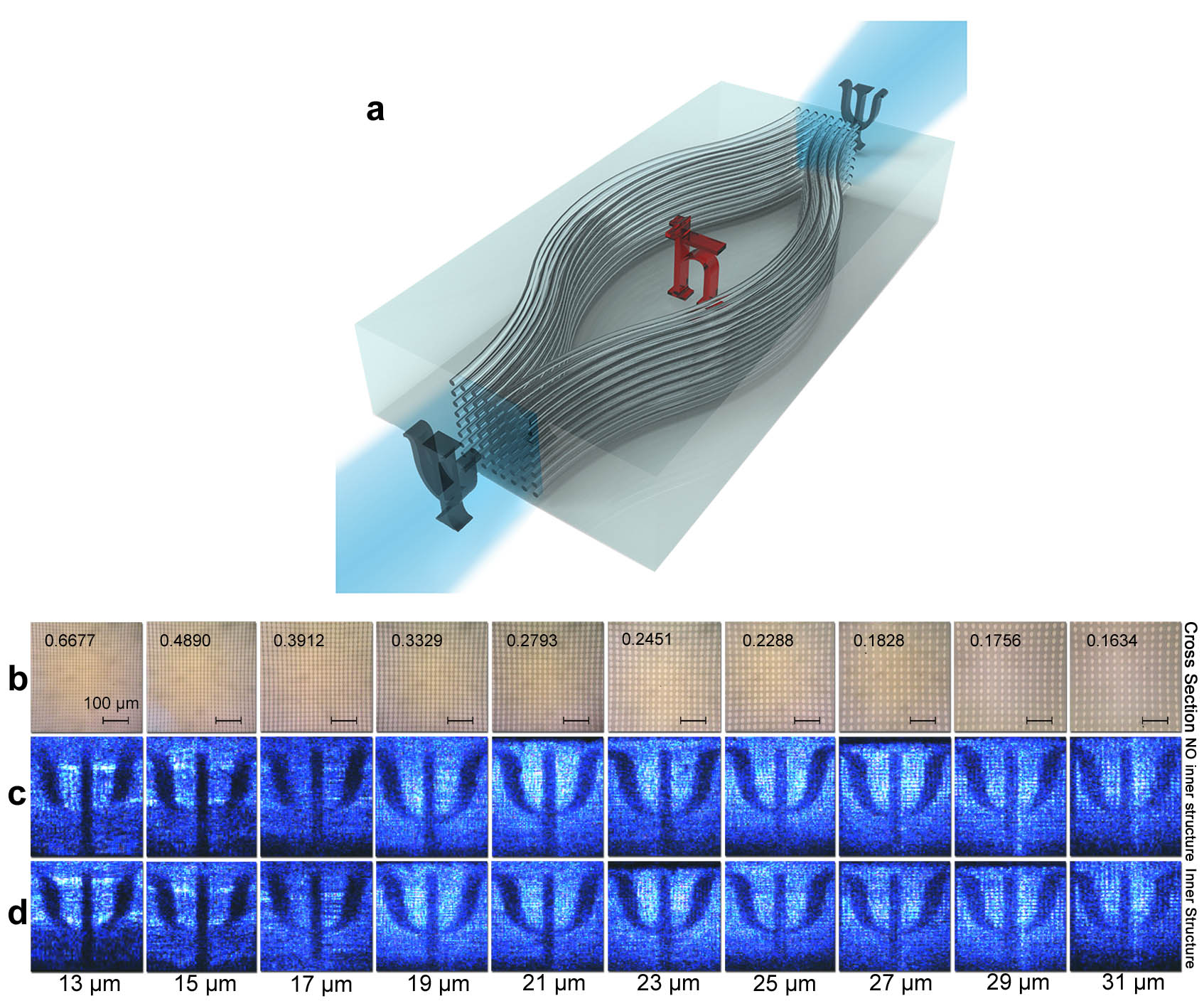}
\caption{\textbf{(a) Principle of on-chip cloaking.} Directional light through a mask of “$\Psi$” is incident onto the window where thousands of waveguides embedded in the chip, propagating along the waveguides to circumvent the inner structure “$\hbar$” and re-assembling to restore the mask image “$\Psi$” without interference of “$\hbar$”. \textbf{(b-d) Characterization of cloaking structures with different pitches:} \textbf{(b)} Cross sections of cloaking windows under the microscope with scale bars and duty cycles. Output images of the cloaking structures with \textbf{(d)} / without \textbf{(c)} the inner structure with pitches varying from $13  \mu m$ to $31 \mu m$. }
\label{fig:figure1}
\end{figure*}
Ray optics provides a new way of concealing macroscopic objects by using calcite crystal \cite{Zhang2011,Chen2011} and standard optical devices such as lens \cite{Choi2014} and prism \cite{Chen2013}, which can guide light though a predefined path by Snell$^\prime$s Law. We introduce a distinctive broadband cloaking scheme in waveguide optics that can be applied on a photonic chip. To be specific, light can be confined in the core of waveguide with higher refractive index associated with a modest curve, which shows its capability of guiding and manipulating light field in an arrayed configuration. Within the tolerance of curvature, a waveguide is bended to circumvent a central region. Optical information through the waveguide is acquired at the output while the exact propagation path remains confidential, which allows hidden objects to exist in the areas rounded by waveguides. The artificial structure requires a waveguide fabrication technique in three-dimensional space, which can be met by the intrinsic processing capacity of femtosecond laser writing  \cite{Davis1996,Nolte2003,Meany2015}. A refractive index contrast can be generated via nonlinear absorption of femtosecond laser in transparent materials, where a trajectory of laser focus transforms to a waveguide. Elaborate control of the trajectory enables complex optical circuits \cite{Crespi2012,Meany2015,Chaboyer2015} as well as three-dimensional configuration \cite{Nolte2003,Chaboyer2015}.
\section*{Results}
As a cloaking scheme, two prerequisites should be met: a clear image transfer and a hidden inner structure \cite{Xu2014}. Foreground mask $\Psi$ featured $0.7mm \times 0.7mm$ is set before the entrance window of the photonic chip and this $\Psi$ image is expected to be soundly transported through the chip without any interference with an inner structure $\hbar$. In order to cloak the $0.44mm \times 0.35mm$ $\hbar$ structure inside a chip, which is fabricated with a high power laser, we prototype up to $5,000$ waveguides in three dimensions along a spindle path opening up the central space for the $\hbar$ to be hidden in (Fig.\ref{fig:figure1}a).\par
In this section, we characterize the cloaking structures by different pitches and optimize the performance with the introduced similarity assessment. Tomography analysis is applied to unveil the evolution of light field in the optimal one. In the end, we benchmark the robustness of the cloaking device with a continuous three-dimensional viewing angle scan.\par
\begin{figure*}[ht]
\centering
\includegraphics[width=0.82\textwidth]{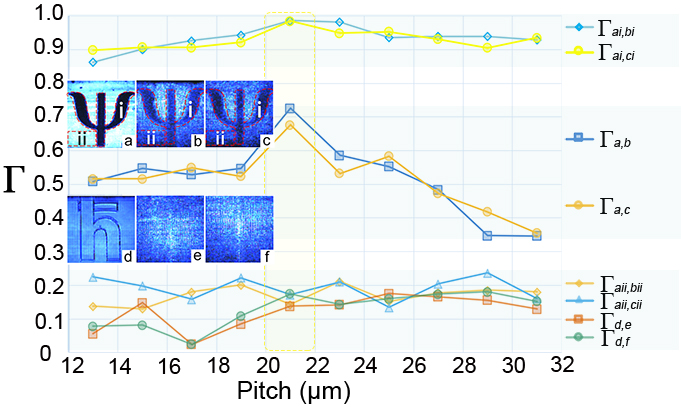}
\caption{\textbf{Area instruction of $\Gamma$ analysis and cloak characterization.} \textbf{(a-c)} show the output images of “$\Psi$”. Area i and ii represent the exact area of “$\Psi$” and the background, respectively: \textbf{(a)} Cloak off and without the inner structure “$\hbar$”. Cloak on and \textbf{(b)} without “$\hbar$” / \textbf{(c)} with “$\hbar$”. \textbf{(d-f)} Gaussian beam without “$\Psi$” mask. \textbf{(d)} Cloak off and with “$\hbar$”. Cloak on and \textbf{(e)} without “$\hbar$” / \textbf{(f)} with “$\hbar$”. \textbf{Cloak characterization.} (1) $\Gamma$ value comparison: $\Psi$ structure ($\Gamma_{ai, bi}$; $\Gamma_{ai, ci}$) $>$ output image ($\Gamma_{a, b}$; $\Gamma_{a, c}$) $>$ background ($\Gamma_{aii, bii}$; $\Gamma_{aii, cii}$) $\approx$ hidden $\hbar$ ($\Gamma_{d, e}$; $\Gamma_{d,f}$). (2) $\Gamma$ values peak at the 21 $\mu m$ pitch of the output image and $\Psi$ structure. (3) The fluctuation of $\Gamma$ values on a low level represents the numerically well cloaked $\hbar$.}
\label{fig:figure2}
\end{figure*}
\subsection*{Cloaking performance optimization}
The spindle of tightly bounded waveguides seems to be a favorable choice to obtain a high resolution image. However, considering the interactions between waveguides, the pitch is obviously a crucial parameter to determine the clarity of image transfer. In our experiment, we fabricate $10$ configurations of spindles with different pitches ranging from $13\mu m$ to $31\mu m$ and the corresponding cross sections captured using a $20\times$ objective lens with a numerical aperture of 0.4 and $16\times$ eyepieces are exhibited in Fig.\ref{fig:figure1}b. A Gaussian beam ($405nm$) is radiated through the mask at the input of the spindle and collected at the output by a CCD camera, as shown in Fig.\ref{fig:figure1}c, d. Each configuration contains two sets: with (Fig.\ref{fig:figure1}d) and without (Fig.\ref{fig:figure1}c) the inner structure.
As illustrated in Fig.\ref{fig:figure1}c and d, various outputs are observed owing to different pitches. In order to evaluate their performances, a uniform statistical criterion should be set up. Therefore we introduce the normalized cross correlation ($\Gamma$) \cite{Lewis1995,Halimeh2011}, a common and convenient assessment of the similarity between two image patterns with the basic idea to compare the grey scale of each corresponding pixel between two images. Specifically, when images are converted into grey matrixes, $\Gamma$ between image \textbf{a} and \textbf{b} can be represented as:
\begin{equation}
\Gamma_{a,b}=\frac{\sum_{i}(a_{i}-\overline{a})(b_{i}-\overline{b})}
{\sqrt{{\sum_{i}(a_{i}-\overline{a})^2 \sum_{i}(b_{i}-\overline{b})^2}}}
\label{equation:1}
\end{equation}
\begin{figure*}[ht!!]
\includegraphics[width=1.8\columnwidth]{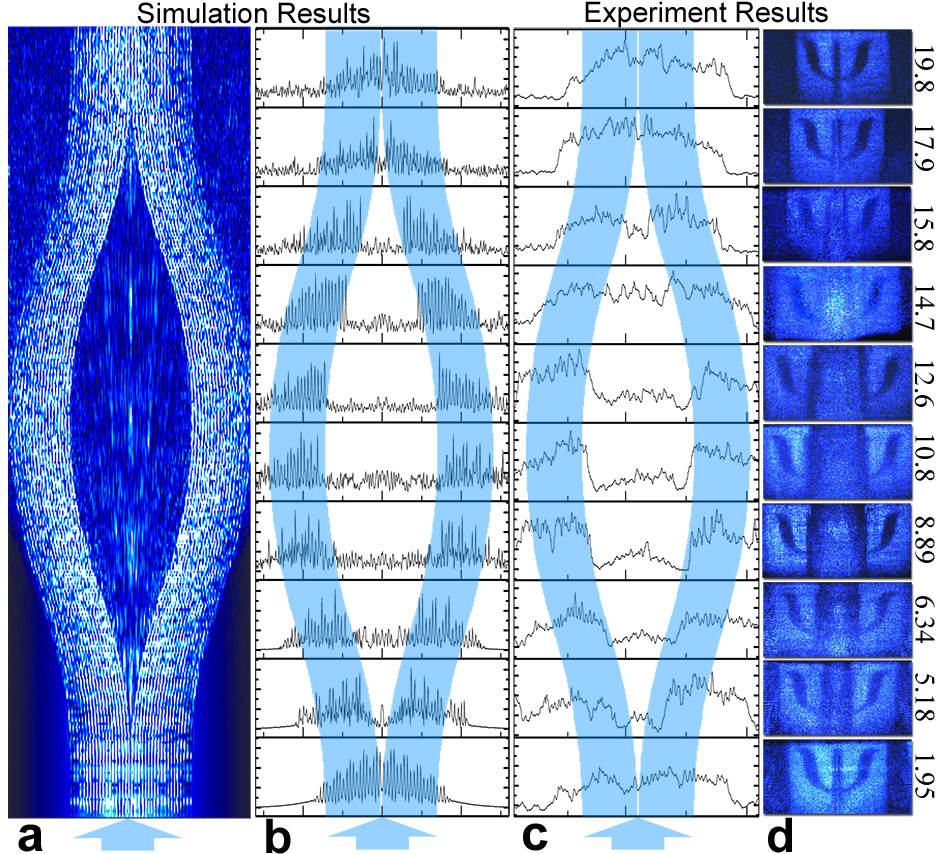}
\centering
\caption{\textbf{Tomography of the cloaking chip.} Simulative \textbf{(a)} and experimental \textbf{(d)} tomography of on-chip cloaking scheme is displayed with intensity distribution along the lateral axis in \textbf{(b)} and \textbf{(c)} respectively. }
\label{fig:figure3}
\end{figure*}
where  $\emph{i}=1,2,\dots$ represents the $i^{th}$ pixel of the image. \emph{$\overline{\textbf{a}}$}, \emph{$\overline{\textbf{b}}$} is the average value of all the grey scale of pixels of \textbf{a} and \textbf{b}, respectively. According to the definition, the $\Gamma$ ranges from $0$ to $1$; the closer to $1$ the value is, the more similar two figures are to each other. \par
\begin{figure*}[ht!!]
\centering
\includegraphics[width=1.5\columnwidth]{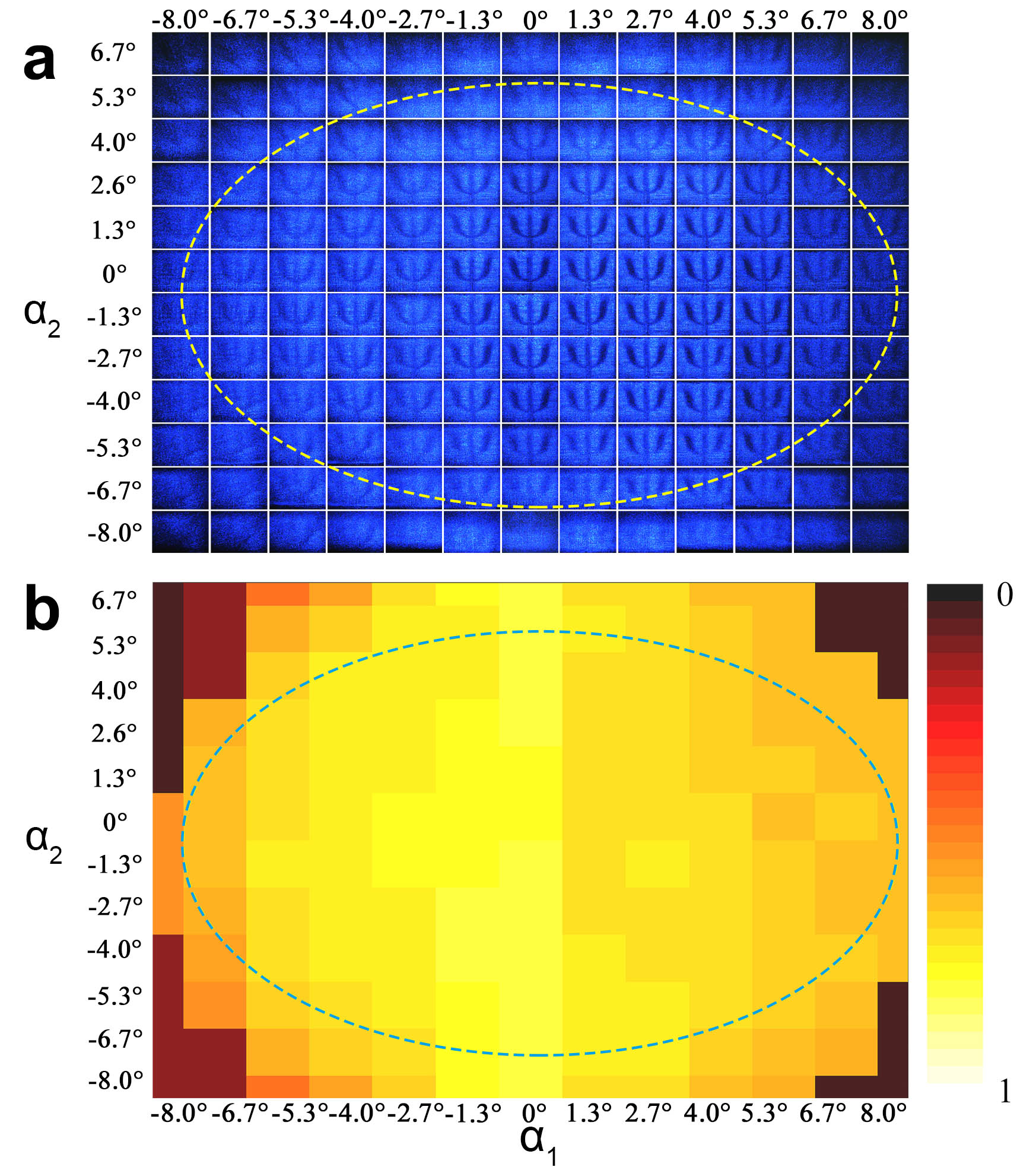}
\caption{\textbf{(a) Image quality on the cloaking chip with different viewing angles.} We get $13 \times 12$ output images by adjusting the lens behind the chip for continuously multi-directional angles and aligning the output image with the corresponding directions. \textbf{(b)Normalized energy angular distribution.} An elliptical-like region spanning $8.0^{\circ}$ laterally and $6.7^{\circ}$ vertically outlines the tolerance for viewing angle in energy distribution.}
\label{fig:figure4}
\end{figure*}
Firstly we compare the similarity of $\Psi$ between the one passing through the waveguides with different pitches and the one only through the chip substrate (Fig.\ref{fig:figure2}b(c) and Fig.\ref{fig:figure2}a). The results ($\Gamma_{a,b}$;$\Gamma_{a,c}$) sketch that the $\Gamma$ value increases steadily and reaches the peak at the pitch value of $21 \mu m$ and then decreases with the pitch getting larger. The optical coupling (crosstalk) distorts the information of the light field propagating in the waveguide when the waveguides get too close to each other. It conforms to a monotonous tendency, which optical coupling decreases by about $85\%$ with every pitch increase of $2\mu m$, according to our calculation. Due to the elliptical cross section (see Fig.\ref{fig:figure1}b), the crosstalk distortion is especially intensive in the vertical direction. On the other hand, a larger pitch constraints the total volume of information in virtue of the lower waveguide density. We calculate the duty cycle, the ratio of the total cross sectional area that the waveguides occupy to the CCD active area, to quantitatively show the waveguide density. (shown in Fig.\ref{fig:figure1}b).\par
As referred in the first result (See line $\Gamma_{a,b}$;$\Gamma_{a,c}$ in Fig.\ref{fig:figure2}), where $21\mu m$ is regarded as the optimal pitch for transmitting the image, the maximum $\Gamma$ value is still moderately large, which can be attributed to the overweight of the background (the region on the image except the $\Psi$) in the calculation of $\Gamma$. To estimate the influence of the background, we extract region ii in Fig.\ref{fig:figure2}a-c and calculate the $\Gamma$, as shown in Fig.\ref{fig:figure2}($\Gamma_{aii,bii}$; $\Gamma_{aii,cii}$). The result shows that the similarity of the background between different images remains at a low level (within the range of $0.18 \pm 0.03$).\par
In view of the low weight of background noise visually, we clip the structure of $\Psi$ (Region i in Fig.\ref{fig:figure2}a-c) from a uniform background and then re-calculated the $\Gamma$. As expected, the $\Gamma$ values increase prominently, reaching the peak at $21 \mu m$ (Fig.\ref{fig:figure2}($\Gamma_{ai, bi}$; $\Gamma_{ai, ci}$)). Furthermore, it is the consequence of the inevitable dissipation of light between waveguides that the maximum of $\Gamma_{ai, bi}$, peaking at $0.986$, still cannot reach $1$. Actually, the pitch of $21 \mu m$ reaches the image standard as high as a retina image.\par
Subsequently, we remove the foreground mask $\Psi$ and compare the similarity between the output image of the chip with the uncloaked inner structure $\hbar$ (Fig.\ref{fig:figure2}d) and the one without the cloaked structure (Fig.\ref{fig:figure2}e), the one with the cloaked structure (Fig.\ref{fig:figure2}f), respectively. The result shows the $\Gamma$ values fluctuate around $0.13 \pm 0.05$, which is numerically hard to differentiate from background noise (see Fig.\ref{fig:figure2} ($\Gamma_{d, e}$; $\Gamma_{d, f}$)). Indeed, the inner structure has been cloaked and cannot be distinguished through calculating the similarity, which obviously accords with our visual result.\par
\subsection*{Tomography analysis of on-chip cloaking}
To reveal the evolution of light field in the printed cloak, tomography analysis is applied to the mechanism of cloaking. We cut the chip transversely, record the data from the output and compare these experiment results with simulation ones. \par
We simulate this experiment by a commercial software RSoft (Beam type: Gaussian, wavelength: $405nm$, refractive index $1.5198$) and derive the distribution of the light field. As shown in Fig.\ref{fig:figure3}, after the beam incidence, most light circumvents the cloaked area along the waveguides, while the rest is scattered by the bulk of the chip and forms the weak white noise in the background. Also we find the light distribution in the output is quite similar with the one in the input, which means little information may have lost after the transmission.\par
Inspired by the widely used tomography in biological science \cite{Fujimoto1995,Low2006}, we perform a tomography on our cloaking chip by grinding the chip to a predefined length, which takes $1.5$ hours for each millimeter removal on a polisher. As the chip gets shorter, by observing the cross section pattern, the process of the projection image splitting into two parts gradually apart from each other and then combining into one again is experimentally reproduced. Comparing the numerical results of the processed intensity distribution along the lateral axis (Fig.\ref{fig:figure3}b), we can observe that the numerical one is more fluctuant, which is because the stray light smooths the fluctuation in the experiment.\\
\begin{figure*}[ht!!]
\centering
\includegraphics[width=0.9\textwidth]{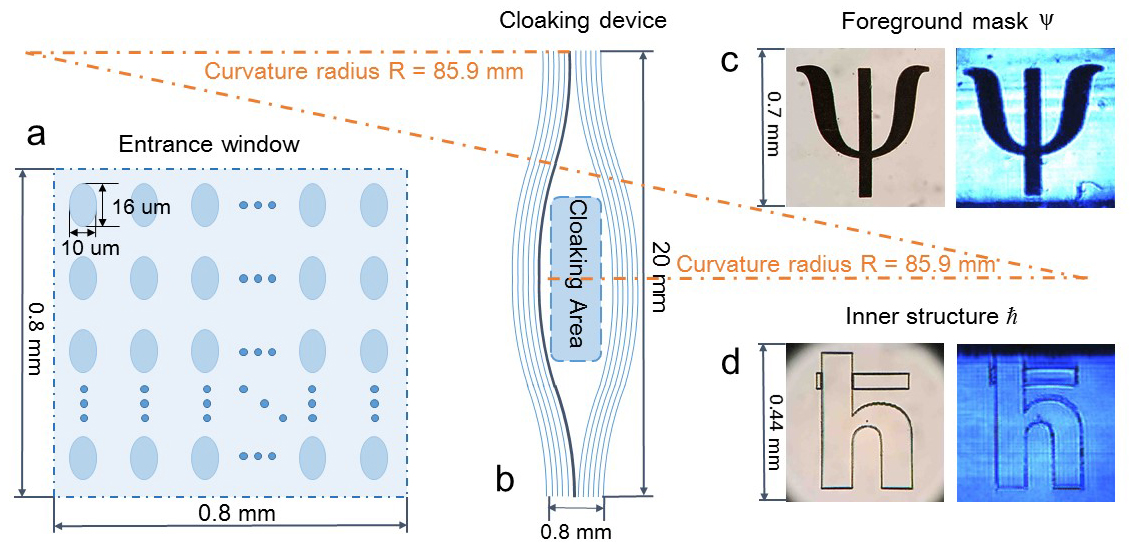}
\caption[width=0.6\textwidth]{\textbf{Overall design parameters.} Design parameters of \textbf{(a)} transverse cross section at both ends, \textbf{(b)} the cloaking structure and \textbf{(c,d)} $\Psi$ mask and \textbf{(d)} inner structure $\hbar$ with their corresponding images.}
\label{fig:figure5}
\end{figure*}
\subsection*{Three-dimensional capacity of cloaking.}
Three-dimensional capacity is frequently employed to benchmark the performance of a cloaking approach. In our experiment, while our cloaking approach is only implemented in one dimension to demonstrate its feasibility, the same method can be extended to three dimensions
with higher complexity of waveguide arrangement and larger number of printed waveguides. We should note that cloaking in one dimension is not that trivial in certain scenarios. For instance, the chips are packaged leaving only some specific access to the public. The components placed at appropriate locations can be well hidden on the chip in the
observation direction. Interestingly, even our one-dimensional configuration has its three-dimensional capacity of cloaking in a cone, which can be understood as an effect of collective numerical aperture of waveguides.\par
Therefore, the tolerance for viewing angle on the chip, which shows three-dimensional capacity of cloaking, is also investigated \cite{Chen2013,Choi2014}. We change the view direction by adjusting the lens behind the cloaking chip ($21\mu m$ pitch, with the inner structure) multi-directionally and align the output image with the corresponding direction as shown in Fig.\ref{fig:figure4}. The sharpness, brightness and visibility decrease as the viewing angle gets large, though the $\Psi$ does not distort. An elliptical-like region spanning $8.0^{\circ}$ laterally and $6.7^{\circ}$ vertically outlines the tolerance for viewing angle. Besides, the limited angle that we derive above conforms to the numerical aperture theory of an individual multimode fiber. It is remarkably interesting that, the inner structure is cloaked so well that even if the obliqueness of the viewing angle is large enough to affect the image transmission, no parts can be observed. In addition, output powers in Fig.\ref{fig:figure4}a are measured and plotted correspondingly (see Fig.\ref{fig:figure4}b) which describes the energy distribution in the output light cone.

\section*{Discussion}
Length difference of waveguides introduces phase distortion site by site. In our experiment, imperfect collective bending of waveguides may generate site-dependent length difference and therefore differential phases. A phase-preserved cloak \cite{Xu2012,Yu2013,Xu2015} can be realized by setting the difference to be integral times of $2\pi$ in the fabrication process of femtosecond direct writing.\par
We further discuss on the capacity of the cloaking structure. In our configuration, we implement the cloaking structure in the curvature radius $85.9mm$ (Definition of raidus in attached supplmentary information) with refractive index contrast $0.0015$ to obtain an optimal cloaking performance in a hidden region in size of $0.44mm \times 0.35mm$. With the decrease of curvature radius, larger bending loss is introduced.\par
In conclusion, we have presented the first experimental demonstration of invisibility cloak in waveguide optics that can hide millimeter-scale structure inside a photonic chip. The printed waveguides confine and guide light field in a three-dimensional array, and evolve with collective curvature in the propagating direction to mimic the Poynting vector in transformation-optics-based cloak \cite{Pendry2006}. A larger invisible volume is attainable with tightly bending cloak spindle of high refractive index contrast \cite{Eaton2011} and rearrangement of waveguides. The quality of image transfer that benchmarks the performance of cloak can be upgraded by increasing the waveguide density and suppressing crosstalk distortion simultaneously, which is achievable if we can manipulate all the phases of evanescent light interference between waveguides \cite{Chaboyer2015}. The on-chip image transfer itself may stimulate novel applications in deep space exploration \cite{Bland2011} and biological sensing \cite{Walt2002}. Furthermore, the printed cloak allows backdoor operations like swap or coupling isolated from observers, representing an emerging problem of information security and hacking on a photopic chip.
\begin{figure*}[ht!!]
\centering
\includegraphics[width=0.75\textwidth]{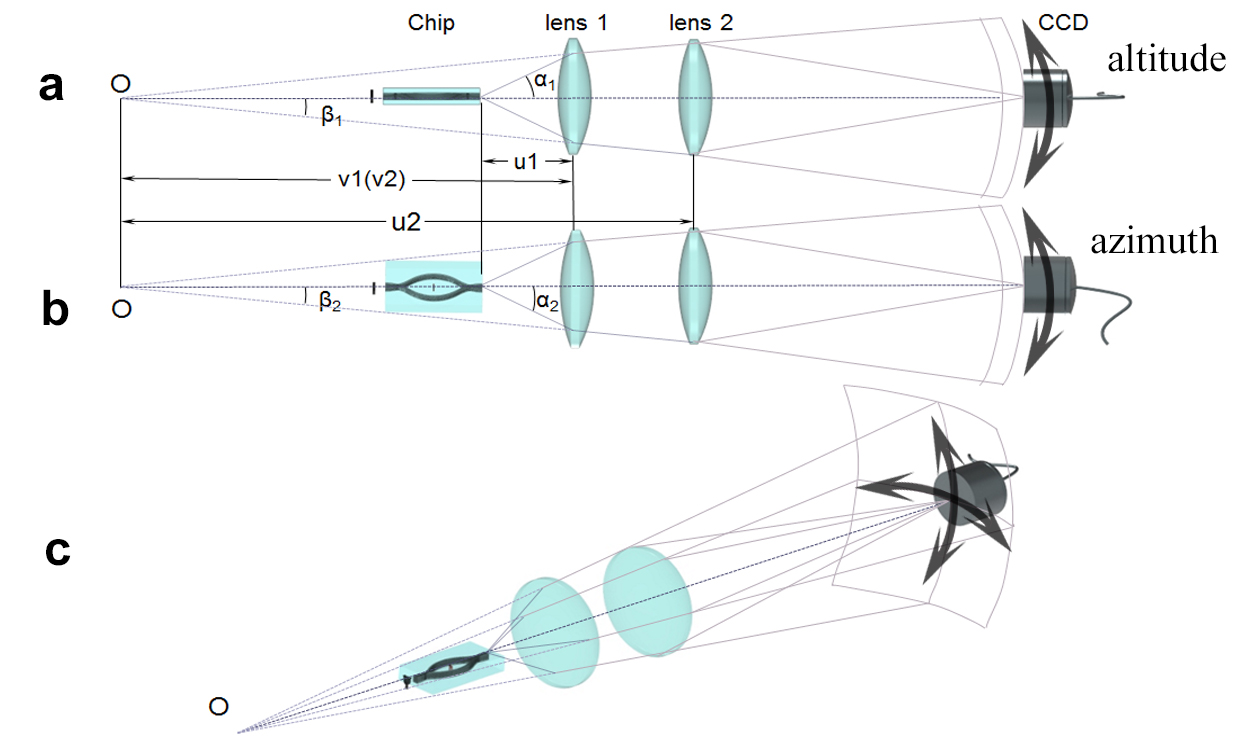}
\caption[width=0.6\textwidth]{\textbf{Schematic of viewing angle measurement. } Viewing angle scan in \textbf{(a)} vertical view, \textbf{(b)} lateral view and \textbf{(c)} 3D view. $u_{i}$ - object distance, $v_{i}$ - image distance. $\alpha_{i}$ can be calculated by the measurement of angle $\beta_{i}(i=1,2)$.}
\label{fig:figure6}
\end{figure*}
\section*{Methods}
\subsection*{Waveguide inscription and laser processing}
A femtosecond laser ($10W,1026nm$) with $290fs$ pulse duration and $1MHz$ repetition rate is frequency doubled to $513nm$ and directed into an spatial light modulator(SLM) to create burst trains which is focused on a borosilicate substrate ($20mm \times 20mm \times 1mm$) with a $50\times$ objective lens with a numerical aperture of 0.55 to fabricate several thousand spindle waveguides ($85.9mm$-radius curves with 1.5mm-long straight design at both ends; refractive index increase of $0.0015$ in the borosilicate substrate $1.5198$) at a constant velocity of $10mm/s$. Array is formed with a cross section size of $0.7mm \times 0.7mm$ and great efforts have been made to process depth independent waveguides through power and SLM compensation. An inner structure of reduced Planck constant $\hbar$ whose visibility is guaranteed by a high power laser direct writing back and forth hundreds of times is located in cloaking area with the size of $0.44mm \times 0.35mm$ (See Fig.\ref{fig:figure5} for overall design parameters).\par
\subsection*{Correlation Coefficient value acquisition}
Since we only concern about the $\Psi$ in images \textbf{a} and \textbf{b} (See Eq.\ref{equation:1}), the exact CCD image clipping is necessary. To eliminate the error caused by the slight shift of $\Psi$, we make a clip of image \textbf{a} with a resolution matrix of $M \times N$, and then crop a smaller box of image \textbf{b} with the matrix of $m \times n (m \leq M, n \leq N)$ in the same way. Transforming the RGB matrix into grey scale, we compare the $m \times n$ matrix extracted from \textbf{b} with every $m \times n$ submatrix of \textbf{a}. The final normalized correlation coefficient value between \textbf{a} and \textbf{b}, $\Gamma_{a, b}$, is chosen to be the maximum of the $(M-m) \times (N-n)$ results. Specifically in our demonstration, we crop image \textbf{a} of $240 \times 320$ pixels and image \textbf{a} of $204 \times 236$ pixels. The $\Gamma_{a, b}$ value is chosen to be the maximum of $(240-204) \times (320-236) = 3024$ results.\par
\subsection*{Simulation of light field evolution}
A commercial Beam Propagation Method software is used to test the cloaking scheme qualitatively in advance. We build a glass block with volume of $20mm \times 1.6mm \times 0.8mm$, transparent boundary condition, refractive index $1.5198$ as a photonic chip where 38$\times$38 waveguides (the size of each cross section is $16\mu m \times 10\mu m$) are imbedded with index increase of 0.0015. All other geometry conforms to the fabrication structure. A $0.8mm \times 0.8mm$ beam, cut from a non-polarized Gaussian beam of $405nm$ wavelength by a square mask, is launched on the cloaking chip. All simulation results are illustrated in Fig.\ref{fig:figure3}a and b.\par
\subsection*{Three-Dimensional Capacity Characterization}
A Gaussian beam (wavelength $405nm$; beam waist $1.8mm$) through the $0.7mm \times 0.7mm$ $\Psi$ mask is radiated at the input of the spindle and  the output light information is collected by two sequent lenses placed behind the chip for imaging, as shown in Fig.\ref{fig:figure6}. The key parameter, the viewing angle $\alpha$ related to the output of the chip cannot be measured directly unless we move the lens 1 after each measurement, which leads to vast workload of readjustment. Since the light is paraxial, we can map the output of the waveguide to the point O by lens 1 and measure the corresponding object distance($u_{1}$) and image distance($v_{1}$). Thus, $\alpha$ can be calculated by the indirect measurement of angle $\beta$ for the relation between $\beta$ and $\alpha$ is given by
\begin{equation}
\alpha_{i} \cdot u_{1}=\beta_{i} \cdot v_{1}, i=1,2
\label{equation:2}
\end{equation}
where $i=1,2$ represents azimuth and elevation angles. Since images can only be formed when lens 2 is within the light cone, we shift lens 2 both in azimuth and altitude direction with a corresponding adjustment of the CCD to catch the image simultaneously. After recording the shifting distance $L_{i}$ (Not shown in Fig.\ref{fig:figure5}) of lens 2 and the length of $u_{2}$, we can get the corresponding angle $\beta_{i}$:
\begin{equation}
\beta_{i} = \frac{L_{i}}{u_{2}}, i=1,2
\label{equation:3}
\end{equation}
Substitute (\ref{equation:3}) into (\ref{equation:2}), we get the expression of viewing angle $\alpha$:
\begin{equation}
\alpha_{i} = \frac{L_{i} \cdot u_{1}}{u_{1} \cdot u_{2}}, i=1,2
\label{equation:4}
\end{equation}
In addition, we have also measured the output powers in every combination of azimuth and elevation angles and plotted in Fig.\ref{fig:figure4}b. This energy distribution demonstrates a radiational decreasing tendency of energy from the center to the edge in the output light cone, which shows a visually great agreement with the radiational decreasing of clearity in the result of Fig.\ref{fig:figure4}a.\par
\noindent\textbf{Acknowledgments:} This research leading to the results reported here was supported by the National Natural Science Foundation of China under Grant No.11374211, the Innovation Program of Shanghai Municipal Education Commission, Shanghai Science and Technology Development Funds, and the open fund from HPCL (No.201511-01). X.-M.J. acknowledges support from the National Young 1000 Talents Plan.

\end{document}